\documentclass[journal]{IEEEtran}
\usepackage{amsthm}
\usepackage{tabu}
\usepackage{lineno}
\usepackage{graphicx}
\usepackage{epstopdf}
\usepackage{latexsym}
\usepackage{lineno}
\usepackage{makecell}
\usepackage{amsfonts}
\usepackage{subfigure}
\usepackage{dsfont}
\usepackage{threeparttable}
\usepackage{multirow}
\usepackage{enumerate}
\usepackage{epsfig}
\usepackage{bm}
\usepackage{cite}
\usepackage{textcomp}
\usepackage{booktabs}  %
\usepackage{xcolor}
\usepackage{amsmath}
\usepackage{algorithm}
\usepackage{algpseudocode}
\newtheorem{remark}{Remark}

\begin{document}

\title{Lightweight 1-D CNN-based Timing Synchronization for OFDM Systems\\with CIR Uncertainty}

\author{Chaojin~Qing,~\IEEEmembership{Member,~IEEE,}
        Shuhai~Tang,
        ~Xi Cai,
        ~and~Jiafan~Wang
\thanks{This work is supported in part by the Sichuan Science and Technology Program (Grant No. 2021JDRC0003), the Major Special Funds of Science and Technology of Sichuan Science and Technology Plan Project (Grant No. 19ZDZX0016 /2019YFG0395), the Special Funds of Industry Development of Sichuan Province (Grant No. 2020-YF09- 00048-SN), and the Demonstration Project of Chengdu Major Science and Technology Application (Grant No. zyf-2018-056).}
\thanks{C. Qing, S. Tang, X. Cai, and J.~Wang are with the School of Electrical Engineering and Electronic Information, Xihua University, Chengdu, 610039, China (E-mails: qingchj@mail.xhu.edu.cn, tangshh@stu.xhu.edu.cn, caixi-1983@163.com, and jifanw@gmail.com).}
}


\maketitle

\begin{abstract}
In this letter, a lightweight one-dimensional convolutional neural network (1-D CNN)-based timing synchronization (TS) method is proposed to reduce the computational complexity and processing delay and hold the timing accuracy in orthogonal frequency division multiplexing (OFDM) systems. Specifically, the TS task is first transformed into a deep learning (DL)-based classification task, and then three iterations of the compressed sensing (CS)-based TS strategy are simplified to form a lightweight network, whose CNN layers are specially designed to highlight the classification features. Besides, to enhance the generalization performance of the proposed method against the channel impulse responses (CIR) uncertainty, the relaxed restriction for propagation delay is exploited to augment the completeness of training data. Numerical results reflect that the proposed 1-D CNN-based TS method effectively improves the TS accuracy, reduces the computational complexity and processing delay, and possesses a good generalization performance against the CIR uncertainty. The source codes of the proposed method are available at https://github.com/qingchj851/CNNTS.
\end{abstract}



\IEEEpeerreviewmaketitle
   \vspace{-2mm}
\section{Introduction}
\IEEEPARstart{O}{rthogonal} frequency division multiplexing (OFDM) has been adopted in many standards\cite{ref:tos}, such as the IEEE 802.11a/n standards, the Internet-of-Things (IoT), and the fifth generation (5G), due to its high spectral efficiency whilst providing resilience to frequency-selective fading \cite{ref:OFDMstd}. In OFDM systems, timing synchronization (TS) plays a key role\cite{ref:tos} in the subsequent channel estimation \cite{ref:JSCE} and symbol detection\cite{ref:JTSDe}. Hence, a variety of classic TS methods have been developed for OFDM systems, e.g., the auto-correlation-based TS \cite{ref:TSsc,ref:RunTS,ref:Auto-Cross} and the cross correlation-based TS \cite{ref:CrossCorr,ref:CrossZC}. However, these classic methods in\cite{ref:TSsc,ref:RunTS,ref:Auto-Cross,ref:CrossCorr,ref:CrossZC} inevitably encounter nonlinear effects, e.g., multi-path interference, which pose a great challenge to improve the accuracy of TS.

To enhance these classic TSs, extreme learning machine (ELM)-based approaches were proposed in \cite{ref:ELM-FTS,ref:myELMTS,ref:ELM-FS}. In \cite{ref:ELM-FTS}, an ELM-based method was proposed to estimate the residual timing offset (TO), in which the coarse TS was assumed to have been accomplished. In \cite{ref:myELMTS} and \cite{ref:ELM-FS}, the ELM networks were employed to refine the TS with preprocessing approaches based on the auto-correlation and cross-correlation, respectively. Although the performance of classic TSs is enhanced by using ELM in \cite{ref:ELM-FTS,ref:myELMTS,ref:ELM-FS} (especially for the scenarios with nonlinear effects in \cite{ref:ELM-FTS,ref:ELM-FS}), several important issues have not been well addressed. Firstly, the ELM network is a single hidden layer-based feed-forward network which leads to limited learning ability. It motivates us to develop a deep learning (DL) network to further enhance the TS accuracy while alleviating substantially computational complexity and processing delay caused by the requirement of preprocessing in \cite{ref:ELM-FTS,ref:myELMTS,ref:ELM-FS}. Secondly, the ELMs in \cite{ref:ELM-FTS,ref:myELMTS,ref:ELM-FS} can be viewed as denoising networks, and the powerful ability of artificial neural network (ANN) has not been well developed, e.g., the feature extraction of TS for OFDM. Finally, these TSs in \cite{ref:ELM-FTS,ref:myELMTS,ref:ELM-FS} are significantly affected by the uncertainty of the power delay profile (PDP) of multi-path channels (or channel impulse response (CIR) uncertainty), due to the fact that the first arriving path is not always the strongest. For example, the efficiency of feature extraction is hindered by using the auto-correlation and cross-correlation as preprocessing, and this degrades the learning ability of ELM-based approaches for the TS in OFDM systems.

Recently, the method to locate the first arriving path is recalled in \cite{ref:OMP} by employing compressed sensing (CS), yet the advantages of ANN were not exploited and the high computational complexity of iteration reconstruction hinders its application. To this end, a high-accuracy ANN-based TS is highly desired to improve the ELM- and CS-based TSs for OFDM systems with a tolerable computational complexity and processing delay in the scenarios of the CIR uncertainty.

To circumvent the above issues, we propose a lightweight one-dimensional convolutional neural network (1-D CNN)-based TS, in which a relaxed restriction of propagation delay would alleviate the CIR uncertainty. To the best of our knowledge, the application of the 1-D CNN and the relaxed restriction of propagation delay is limited, while our solution fills this gap by improving the TS accuracy against the CIR uncertainty with the reduced computational complexity and processing delay (i.e., network lightweight).
The main contributions are summarized as follows:
\begin{itemize}
  \item
    We propose a 1-D CNN-based method to improve the TS accuracy from the perspective of classification tasks. In this developed method, the TS is transformed into a DL-based classification problem. In particular, the advantage of CNN in performing the feature extraction is exploited, and thus highlights its classification features for TS. Compared with TS methods in \cite{ref:myELMTS,ref:ELM-FTS,ref:ELM-FS}, the TS accuracy is significantly improved.
  \item
We develop a lightweight NN structure to reduce the processing delay and computational complexity. Especially, the CNN architecture is simplified by unfolding a few iterations of CS-based TS in \cite{ref:OMP}. Besides, the dedicated filter size and filter number of each CNN layer are designed to highlight the classification features (from the receptive fields of TS features) without increasing the parameters of the network. In this way, features of the training sequence are effectively extracted with a relatively low processing delay and computational complexity.

  \item
  We exploit the relaxed restriction of propagation delay to augment the completeness of training data for alleviating the CIR uncertainty.
  Particularly, a propagation delay that is greater than the real one and a random PDP are exploited to increase the completeness of training data. This relaxed method improves the generalization performance of the proposed 1-D CNN-based TS against the CIR uncertainty, and thus possesses a good generalization capacity compared with those TS methods in \cite{ref:myELMTS,ref:ELM-FTS,ref:ELM-FS}.
\end{itemize}


\textit{Notations:} The notations adopted in this letter are described as follows. $\mathbb{R}^{M\times N}$ and $\mathbb{C}^{M\times N}$ stand for the $M$-by-$N$ dimensional real matrix space and $M$-by-$N$ dimensional complex matrix space, respectively. $[\cdot]^T$, $\mathrm{E}\{\cdot\}$,  $|\cdot|$, $\lceil\cdot\rceil$, and $\lfloor\cdot\rfloor$stand for the transpose operation, expectation operation, absolute operation, ceiling function, and floor function, respectively. $[\mathbf{X}]_{m,n}$ denotes the entry $(m,n)$ of matrix $\mathbf{X}$, respectively. $\Re(x)$ and $\Im(x)$ denote the real and imaginary parts of value $x$, respectively.
\begin{figure}[t]
  \centering
  \vspace{-2mm}
  \includegraphics[width=0.44\textwidth]{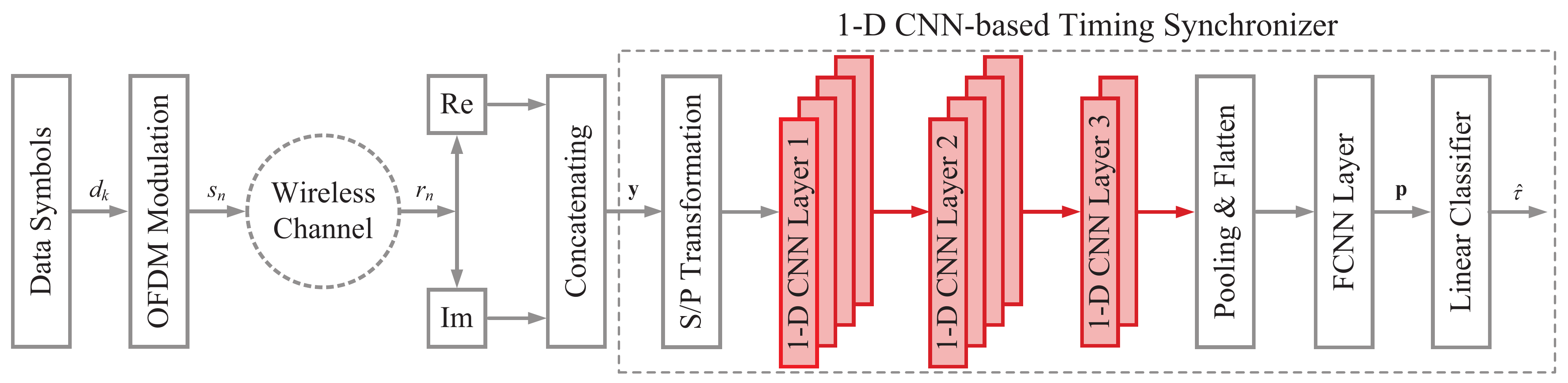}\\
  \caption{System model and 1-D CNN-based timing synchronizer.}\label{figSM}

\end{figure}
   \vspace{-2mm}
\section{System Model}

{The system model is given in Fig.~\ref{figSM}, in which the OFDM modulation is employed.}
{In the frequency domain,} the data symbol $d_k$, with $0\le k\le N-1$ and $N$ being the number of sub-carriers, is {transformed} to the time domain by {using} inverse discrete Fourier transform (IDFT){, and its $n$-th sample (denoted as $s_n$) is given by}
\begin{equation}\label{EQ: FtoT}
{s_n} = \frac{1}{{N}}\sum\limits_{k = 0}^{N - 1} {{d_k}\exp \left( j2\pi{\frac{{ kn}}{N}} \right)},
\end{equation}
where $s_n$ satisfies that $\mathrm{E}\{|s_n|^2\}=\sigma^2_d$, with $\sigma^2_d$ being transmitted signal power.
After appending the $L_c$-length cyclic prefix (CP) and transmitting OFDM signals over the wireless channel\cite{ref:CHMod}, the $n$th sample of received signal is expressed as:
\begin{equation}\label{EQ:Rxdata}
{r_n} = \sum\limits_{p = 1}^{P} {{h_{{\tau}_p}}{s_{n - \widetilde{\tau}  - {\tau}_p}}} \exp \left( {j2\pi \varepsilon \frac{{{n - \widetilde{\tau} }}}{N}} \right) + {w_n},
\end{equation}
where $P$, $h_{{\tau}_p}$, $\widetilde{\tau}$, ${{\tau}_p}$, and $\varepsilon$ separately stand for the number of resolvable multi-path, the channel complex gain of $p$-th path, the unknown TO to be estimated, the delay of $p$-th path relative to the first arriving path, and carrier frequency offset (CFO). $w_n$ denotes the circularly symmetric complex Gaussian noise, {i.e.,  ${w_n} \sim \mathcal{CN}(0, \sigma^2_n)$ with $\sigma^2_n$ being the noise power.} By extracting the real and imaginary parts of the $M$-length received signal, the input of 1-D CNN-based timing synchronizer, i.e, $\textbf{y}\in\mathbb{R}^{2M\times1}$ in Fig.~\ref{figSM}, is formed as
\begin{equation}\label{EQ:yinput}
{\bf{y}} = {\left[ {\Re(r_0),\Im(r_0), \cdots , \Re(r_{M - 1}),\Im(r_{M - 1})} \right]^T},
\end{equation}
where $M$ ($M=N_u+N$) is the length of observation interval and $N_u=N+L_c$.
In this letter, we design a lightweight DL-based timing synchronizer to estimate the TO $\widetilde{\tau}$ against the CIR uncertainty.
 \vspace{-1em}
\section{1-D CNN-based Timing Synchronizer}
In this section, the proposed 1-D CNN-based TS for OFDM systems is elaborated. In Section III-A, the architecture of the proposed timing synchronizer is characterized in detail and then followed by the relaxed restriction for propagation delay in Section III-B. Next, Section III-C presents the offline and online implementations. Finally, the computational complexity and processing delay are given in Section III-D.
\vspace{-1em}
\subsection{Lightweight Network Architecture}
To reduce the computational complexity and processing delay, a lightweight network for TS is highly desired. To this end, we propose a 1-DCNN-based timing synchronizer and make a dedicated design for its filter size. The network architecture is summarized in Table~\ref{table_I} (its processing procedure is presented in Fig. \ref{figSM}), and the detailed descriptions are given as follows.

This timing synchronizer consists of a three-layer 1-D CNN along with a single-hidden-layer fully connected neural network (FCNN) layer. In the proposed synchronizer, the first CNN layer is related to the processing of the classic cross-correlation, and the second and the third CNN layers aim to extract the significant features of the training sequence, followed by the FCNN layer to make the final decision about the TO estimation. In these three CNN layers, the valid padding and stride 2 are adopted for the first CNN layer to reap the real and imaginary parts of $r_n$ given in \eqref{EQ:yinput}, $0\le n\le M-1$, while the same padding and stride 1 are applied for the rest CNN layers to retain significant features.
Each CNN layer is activated by a rectified linear units (ReLU) activation function. For the FCNN layer, one hidden layer and the tanh activation function are employed to fuse TS features. Between CNN layers and the FCNN layer, an average pooling layer is utilized to reduce the computational complexity. At the output layer, a softmax activation function is adopted, and is followed by a linear classifier to produce the candidates of TO.


\begin{table}
 \vspace{-1em}
\renewcommand{\arraystretch}{1.1}
\caption{Network Architecture of 1-D CNN-based Timing Synchronizer}
\label{table_I}
\centering
\tiny
\setlength{\tabcolsep}{1.25mm}{
\begin{tabu}{c|c|c|c|c}
\tabucline[0.75pt]{-}
    Layer Name       & Size of output  &  Size of filter &Number of filter & Activation\\ \tabucline[0.75pt]{-}
    Input Layer      & $2M\times1\times1$  & - &-&-\\ \hline
     CNN Layer 1  & $N_u\times1\times4$  & $(2N+1)\times1$ &4& ReLU \\ \hline
     CNN Layer 2  & $N_u\times1\times4$  & $(\lceil L_c/2\rceil+1)\times1$ &4& ReLU \\ \hline
     CNN Layer 3  & $N_u\times1\times2$  & $(\lceil L_c/2\rceil+1)\times1$ &2& ReLU \\ \hline
    Average pooling  & $\lfloor N_u/2\rfloor \times1\times2$  & - & - & -\\ \hline
    Flattening  & $2\lfloor N_u/2\rfloor\times1$  & -&  -& -\\ \hline
    FCNN Layer  & $N_u\times1$  & -& - & Tanh\\ \hline
    Output Layer & $N_u\times1$  & -& -  &Softmax\\
    \tabucline[0.75pt]{-}
\end{tabu}}
\vspace{-1mm}
\end{table}
\vspace{-0.5em}
\begin{remark}
 The lightweight network is mainly reflected in the three-layer 1-D CNN architecture. Based on the model-driven mode, we unfold and simplify three iterations of CS-based TS in \cite{ref:OMP} to form three simplified CNN layers (the simplified details are given in \textbf{Remark 3.}). Since only three iterations are unfolded and these iterations are simplified, the proposed method is extremely lightweight, so that its computational complexity and processing delay are significantly reduced compared with \cite{ref:OMP}. Besides, the 1-D CNN is developed rather than the 2-D CNN (or FCNN), and the pooling layer is also employed, leading to the network being lightweight as well.
\end{remark}
\vspace{-0.5em}
\begin{remark}
Usually, the TS accuracy is closely related to the timing metric (TM). In this letter, we improve the TM by using a dedicated design of filter size.
First, the ($2N+1$)-length filter size is utilized to form a relatively large receptive field for capturing the complete features of TM in the first CNN layer. The main consideration is to involve the whole features of the training sequence to replace and enhance the preprocessing of ELM-based TS in \cite{ref:ELM-FTS,ref:myELMTS,ref:ELM-FS}, e.g., the cross-correlation-based preprocessing in classic TSs. This is different from the ELM-based TS in \cite{ref:ELM-FTS,ref:myELMTS,ref:ELM-FS} since the preprocessing is not adopted.
Then, the relatively small filter size, i.e., $\lceil L_c/2\rceil+1$ ($L_c$ denotes the length of CP), is developed for both the second and third CNN layers to extract and refine the significant features of TM. Due to the relatively small filter size, the processing delay and computational complexity are reduced.
\end{remark}
\vspace{-0.5em}
\begin{remark}
For simplifying CNN layers, we also develop a relatively small number of filters and make a trade-off among the three layers. In the first CNN layer, four filters are adopted to maintain the complete features of TM, which are held and highlighted by another four filters employed in the second CNN layer. In the third CNN layer, we only use two filters to refine the significant features of TM. By this means, high relevant features of the training sequence are efficiently extracted, thus facilitates the subsequent classification in the FCNN layer. Also, the number of filters in each CNN layer is small and the convolutional process is the linear weighted sum of two 1-D data so that CNN layer is simplified.
\end{remark}
\vspace{-2em}
\subsection{Relaxed Restriction for Propagation Delay}
Although the ELM-based TSs in \cite{ref:ELM-FTS,ref:myELMTS,ref:ELM-FS} improve the TS accuracy of classic TS methods to some extent, they inevitably encounter the CIR uncertainty. The CIR uncertainty weakens the PDP correlation between the training CIRs and the real ones, and thus degrades the learning efficiency of NN in ELM-based TS. To tackle this challenge, we relax the restriction for PDP of CIRs by extending its maximum propagation delay to augment the completeness of training data (especially for the scenarios that have not been experienced). The received power in the scenarios of line of sight (LOS) is $P_r = \frac{\lambda^2}{(4\pi \zeta)^{2}}\sigma^2_d G_t G_r$ \cite{ref:LoSloss} with $\lambda$, $\zeta$, $G_t$, and $G_r$ being the wavelength, propagation distance, transmitted antenna gain, and received antenna gain, respectively. Then, the maximum propagation delay {$\tau_P$} in non-line-of-sight (NLOS) scenarios is given by
{\begin{equation}\label{EQ:tauMAX}
\tau_P = \frac{\zeta_{max}} c - |\Delta \tau|,
\end{equation}
where $c$ denotes the speed of light, $|\Delta \tau|$ is the delay fluctuation, and $\zeta_{max}$ stands for the maximum propagation distance (resolvable power $P_r$) with given transmitted power $\sigma^2_d$ in LOS scenarios. We employ the maximum propagation distance ${\zeta_{max}}$ in LOS with given $\sigma^2_d$ as the relaxed restriction for NLOS, since its propagation distance is inevitably reduced by encountering obstacles.
By denoting $\tau_{\rm relax}$ as the relaxed restriction for propagation delay, we have
\begin{equation}\label{EQ:RELAX_MAX}
\tau_{\rm relax}= \frac{\zeta_{max}} c = {\tau_P} + |\Delta \tau|.
\end{equation}
With the relaxed restriction of propagation delay $\tau_{\rm relax}$, the channel model that we employ to construct the training data set obeys the exponentially decayed PDP \cite{ref:CHMod} with its exponent ${\eta}$ satisfying ${\eta}\sim\mathcal{U}(0.01,0.5)$.
Although the channels with exponentially decayed PDP are employed for training, the trained network can effectively work in most channel scenarios, e.g., the 5G channel models given in 3GPP TR 38.901 \cite{ref:3GPP5G}, possessing excellent generalization performance.

\begin{remark}
According to the above $\tau_{\rm relax}$ and $\eta$, the completeness of training data and the robustness of timing \cite{ref:myELMTS} have been improved. In other words, a larger maximum multi-path delay and a random PDP may generate real CIR profiles with higher probability, so that the generalization performance of NN in our TS method is boosted against the CIR uncertainty.
\end{remark}
\vspace{-1em}
\subsection{Offline Training and Online Deployment}
The DL framework\cite{ketkar2017deep} named Keras functional API along with {TensorFlow} is employed to implement the 1-D CNN-based timing synchronizer.
The processes of offline training and online deployment are given as follows, respectively.

\subsubsection{Offline Training}

For training data, the relaxed restriction for propagation delay, i.e., $\tau_{\rm relax}$, and $\eta$ in Section III-B are employed.  {$N=128$ ($L_c=N/4=32$ \cite{ref:CPLength}) and $P=23$ \cite{ref:3GPP5G} ($\tau_P$ uses $(P-1)$ sample periods\cite{ref:ch_pdp}, i.e., $\tau_P = 22$) are considered.} The training data set $\bm {\aleph}$ with $N_t$ ($N_t=10^5$) training samples is expressed as
\begin{equation}\label{data_set}
{\bm \aleph}=\{\mathbf{r}_i,\mathbf{t}_i\}^{N_t}_{i=1},
\end{equation}
where $\mathbf{r}_i\in\mathbb{C}^{M\times1}$ is the training data with its entries given in \eqref{EQ:Rxdata}, and $\mathbf{t}_i\in\mathbb{R}^{N_u\times1}$ denotes the corresponding TS label. According to the one-hot coding \cite{ref:myELMTS}, $\mathbf{t}_i$ is given by
\begin{equation}\label{data_label}
\small
{{\bf{t}}_i} = {\left[ {\underbrace {0 \cdots 0}_{\left\lceil {\widetilde{\tau} _i}+{\frac{1}{2}\left( { {L_c} + {\tau _{{\rm{relax}}}}} \right)} \right\rceil }1\underbrace {0 \cdots 0}_{{N_u} - \left\lceil {\widetilde{\tau} _i}+{\frac{1}{2}\left( { {L_c} + {\tau _{{\rm{relax}}}}} \right)} \right\rceil  - 1}} \right]^T},
\end{equation}
where $\widetilde{\tau}_i$ is the TO of the $i$-th training data,$\widetilde{\tau}_i \sim \mathcal{U}(0,N-1)$,  {$\Delta\tau=\lceil(L_c-\tau_P)/2\rceil = 5$, and $\tau_{\rm relax}=\tau_P + |\Delta \tau|=27$ sample periods according to (5)} for the offline training phase.



According to \eqref{EQ:yinput}, the model input $\mathbf{y}_i\in\mathbb{R}^{2M\times1}$ is obtained. Since the TS is transformed into the DL-based classification problem, its model output $\mathbf{p}_i\in\mathbb{R}^{N_u\times1}$ corresponds to the probability of candidate TO, which is given by
\begin{equation}\label{PROB_COMPUT}
p_{i,j} = \frac{\exp(o_{i,j})}{\sum\nolimits_{j = 0}^{{N_u} - 1} {\exp \left( {{o_{i,j}}} \right)} },
\end{equation}
where $o_{i,j}=[\textbf{o}_i]_{j,1}$ and $p_{i,j}=[\textbf{p}_i]_{j,1}$, with $\textbf{o}_i$ being the input of the last layer of FCNN, and $\mathbf{p}_i$ being the output of FCNN.

The cross entropy ($\sum\nolimits_{j = 0}^{{N_u} - 1} {{t_{i,j}}\log {p_{i,j}}}$ with $t_{i,j}=[\mathbf{t}_i]_{j,1}$) is adopted as the loss function, in which the stochastic gradient descent (SGD) with Adam optimizer is employed. The optimizer parameters are set as $\alpha=0.001$, $\beta_1=0.9$, and $\beta_2=0.999$\cite{kingma2014adam}. During the training phase, the DL-based model $f_m$ is fitted with ${\bm \aleph}$, which is expressed as
\begin{equation}\label{Lear_fit}
{f_m} \leftarrow \mathrm{Train}{_{{f_m}}}\left( {{\bm \aleph},{\bf{\Theta}},e} \right),
\end{equation}
where ${\bf \Theta}$ denotes the network parameters to be optimized, and $e$ represents the $e$-th training processes that contains the usual operations of forward (back) propagation and weights update \cite{kingma2014adam}. Since the TS is to locate the TO in the ISI-free region, the trained network parameters is formulated as
\begin{equation}
{\bf \Theta}  \leftarrow \mathop {\min }\limits_{{R_{\textrm{err},e}}}\{ \mathrm{Train}{_{{f_m}}}\left( {{\bm \aleph},{\bf \Theta},e} \right)\},
\end{equation}
where $R_{\textrm{err},e}= [\sum\nolimits_{i = 1}^{{N_t}} {{\theta\left(\widehat{\tau}_i,\widetilde{\tau}_i\right)}}]/{{{N_t}}}$ is utilized to evaluate the accuracy of TS and save the best network parameters after the $e$-th training process, and $ \theta \left( {{\widehat{\tau _i}},{\widetilde{\tau} _i}} \right)$ is given by
\begin{equation}\label{EQ: ThetaFunc}
\theta \left( {{\widehat{\tau _i}},{\widetilde{\tau} _i}} \right) = \left\{ \begin{array}{l}
0,\ 0 \le {\widetilde{\tau} _i} - {\widehat{\tau} _i} \le L_c-{\tau _{{\rm{relax}}}}\\
1,\ {\rm{others}}
\end{array} \right.,
\end{equation}
where the estimated TO is expressed as
\begin{equation}\label{EQ: ESTSTO}
\widehat{\tau _i}=\mathop {\arg \max }\limits_{0 \le j \le {N_u} - 1} \{p_{i,j}\}.
\end{equation}

\subsubsection{Online Deployment}
With the trained 1-D CNN-based timing synchronizer, its online running performs as the TS for OFDM systems. By using \eqref{EQ: FtoT}--\eqref{EQ:yinput}, the model input of timing synchronizer, i.e., $\mathbf{y}\in\mathbb{R}^{2M\times1}$, is obtained. With the input $\mathbf{y}$, $\{p_j\}^{N_u-1}_{j=0}$ is produced, and thus the TO estimation is
\begin{equation}\label{EQ: ESTSTOonline}
    \widehat{\tau}=\mathop {\arg \max }\limits_{0 \le j \le {N_u} - 1} \{p_j\}.
\end{equation}
\vspace{-2em}
\subsection{{Computational Complexity and Processing Delay}}
The computational complexity and processing delay among different methods for TS are elaborated in Table \ref{table_II}, in which the complex multiplication (CM) is employed to as the metric of computational complexity. $N_l$ and $N_{l-1}$ denote the neuron numbers of the current and previous layers, respectively; $K_l$ is the filter size of the $l$-th CNN layer; $C_l$ and $C_{l-1}$ represent the filter numbers of current and previous CNN layers, respectively; $\mathcal{L}_c$ and $\mathcal{L}_d$ stand for the total layer numbers of 1-D CNN and FCNN layers, respectively. For the CS-based TS method in \cite{ref:OMP}, the constraint length orthogonal matching pursuit algorithm is employed as the comparison. For a fair comparison, the complete synchronization process in \cite{ref:OMP} is selected to compute its CMs and processing delays, since the method of \cite{ref:OMP} cannot complete the TS without estimating or equalizing CIRs.
Due to the lightweight network architecture, the CM of the proposed method is significantly lower than those of the CS- and ELM-based methods (in \cite{ref:myELMTS} and \cite{ref:OMP}, respectively). This is verified by the example given in Table \ref{table_II}, where $N=128$, $L_c=32$, $N_u=160$, and $P=28$ are considered. Meanwhile, $10^4$ simulations are employed to evaluate the processing delay, and the proposed 1-D CNN-based TS method reaps the smallest processing delay among different TS methods.
On the whole, compared with the CS and ELM-based TSs, the proposed 1-D CNN-based TS reduces the processing delay and computational complexity due to the lightweight NN architecture.
\begin{table}
\vspace{-1em}
\renewcommand{\arraystretch}{1.1}
\caption{Computational Complexity and Processing Delay among Different TS Methods}
\label{table_II}
\centering
\tiny
\setlength{\tabcolsep}{1.5mm}{
\begin{tabu}{c|c|c|c}
\tabucline[0.75 pt]{-}
 Method      & Computational Complexity  &CM& Time (sec)\\
 \tabucline[0.75 pt]{-}
    CS \cite{ref:OMP}      & $\mathcal{O}[PN{N_u} + \sum\nolimits_{p = 1}^P ({3p{N_u}}  + {p^3} + {p^2}{N_u})]$              &  $2167396$  & $81.41$            \\ \hline
    ELM \cite{ref:myELMTS}   & $\mathcal{O}[\frac{3}{2}N+4(N_u-1)+16N_u^2]$              &  $410428$ & $1.61$        \\ \hline
    {Proposed}      & $\mathcal{O}[\frac{1}{4}(\sum\nolimits_{l=2}^{\mathcal{L}_c} {{N_l}{K_l}{C_l}{C_{l - 1}}}  +\sum\nolimits_{l=2}^{\mathcal{L}_d}{{N_l}{N_{l - 1}}})]$              &  ${70240}$   & $0.35$           \\
    \tabucline[0.75 pt]{-}
\end{tabu}}
\vspace{-2mm}
\end{table}
\begin{figure*}[t]
  \centering
  \vspace{-3mm}
   \includegraphics[width=0.95\textwidth]{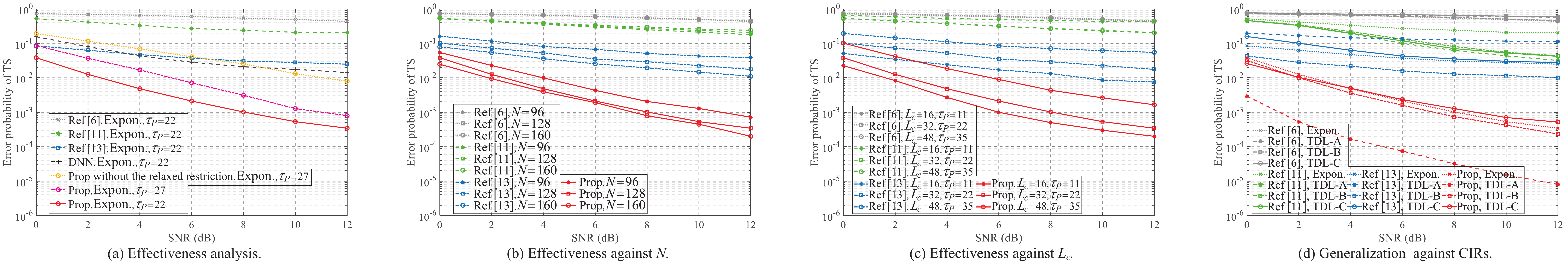}\\
      \vspace{-2mm}
  \caption{The error probability of TS versus SNR.}\label{figEff}
     \vspace{-4.5mm}
\end{figure*}
\vspace{-0.25em}
\section{Numerical Results}
This section provides numerical results to evaluate the synchronization performance. Basic parameters involved in the simulation are given as follows.
 {$N=128$, $L_c=N/4=32$\cite{ref:CPLength}, $N_u=N+L_c=160$, and $M=N_u+N=288$ are considered.}
The transmitting training sequence for TS is the Zadoff-Chu sequence\cite{ref:ZCseq}. For the convenience of expression, we leverage the ``Prop'', ``Ref \cite{ref:OMP}'', ``Ref \cite{ref:myELMTS}'', and ``Ref \cite{ref:RunTS}'' to denote the proposed TS method, the CS-based TS in \cite{ref:OMP}, the ELM-based TS in \cite{ref:myELMTS}, and the classic TS method in \cite{ref:RunTS}, respectively.
Besides, the ``DNN'' denotes a four-hidden-layers DNN-based TS method that has a similar computational complexity as the ``Prop''.
\vspace{-1em}
\subsection{Effectiveness Analysis}
To analyze the effectiveness of the proposed TS method, Fig.~\ref{figEff}(a) depicts the comparison of error probabilities,
where $P=23$, $\eta=\frac{1}{P}=\frac{1}{23}$ \cite{ref:ch_pdp}, and ${\tau}_{P}$ is 22 sample periods for testing.
For each given signal-to-noise ratio (SNR) in Fig.~\ref{figEff}(a), the ``Prop''  reaches the smallest error probability of TS among the given TS methods. Especially, with the increase of SNR, the ``Prop''  remarkably reduces the error probability compared with other TS methods.
For example, all the error probabilities of ``Ref \cite{ref:OMP}'', ``Ref \cite{ref:myELMTS}'', and ``Ref \cite{ref:RunTS}'' are higher than $2\times10^{-2}$ in whole given SNR region, while the error probabilities of the ``Prop''  are lower than $10^{-2}$ when SNR$\geq3$dB.
Also, the error probability of the ``Prop''  is lower than that of the ``DNN'' for each given SNR since its complex parameter tuning results in
the difficulty of optimizing the network parameters.
Besides, compared with the  {``Prop without the relaxed restriction''}, the ``Prop''  achieves a less performance degradation when testing ${\tau}_P$ increases from 22 to 27. This validates the effectiveness of using the relaxed restriction of propagation delay.
Meanwhile, Fig.~\ref{figEff}(b) and Fig.~\ref{figEff}(c) depict the error probabilities of TS against the number of sub-carriers and the length of CP, respectively. Except for those parameters (presented in Fig.~\ref{figEff}), other parameters are the same as those in Fig.~\ref{figEff}(a). For each given SNR in Fig.~\ref{figEff}(b) and Fig.~\ref{figEff}(c), the ``Prop'' achieves the lowest error probability given different values of $N$ and $L_c$, respectively.
To sum up, the proposed 1-D CNN-based TS method (i.e., ``Prop'') can effectively reduce the TS's error probability with the reduced complexity and processing delay.
\vspace{-1.25mm}
\subsection{Generalization Analysis}
To analyze the generalization performance of the proposed method against the CIR uncertainty, Fig.~\ref{figEff}(d) plots the error probability of TS, where the PDPs of TDL-A, TDL-B, and TDL-C given in 3GPP TR 38.901~\cite{ref:3GPP5G} are employed for comparison. {It is noteworthy that  the relaxed restriction for propagation delay in Section III-B is adopted for training the ``Prop'', while the testing PDPs of  TDL-A, TDL-B, and TDL-C are acquired by setting sample periods ${\tau}_P$ as 22, 22, and 23, respectively.}
According to Fig.~\ref{figEff}(d), the ``Prop''  achieves the smallest error probability among the given TS methods in each given channel scenario.
This is because the proposed relaxed restriction increases the completeness of training data, improving the generalization performance of NN against the variation of CIRs.
For example, when SNRs are greater than $4$dB, the maximum error probability of the ``Prop''  is smaller than $8\times10^{-4}$ for each given channel scenario, while the error probabilities of the other TS methods are greater than $2\times10^{-3}$.
Especially for the case where the testing channel scenario is TDL-A, the ``Prop''  achieves significant performance improvement in reducing the error probability.
According to empirical simulation results, the PDP of TDL-A channel is close to the PDPs of training channels, and therefore the ``Prop''  performs well under TDL-A channel.
As a result, the proposed 1-D CNN-based timing synchronizer combined with the proposed relaxed restriction for propagation delay possesses excellent generalization performance against the CIR uncertainty.
   \vspace{-1.75mm}
\section{Conclusion}
In this letter, a lightweight 1-D CNN-based TS for OFDM is investigated to alleviate the CIR uncertainty. The developed timing synchronizer highlights the TS features and thus improves the TS accuracy from the perspective of DL-based classification tasks. Especially, by carefully designing the filter size, the number of filters, the network architecture in our timing synchronizer is simplified, and the computational complexity and processing delay are reduced compared with those in the CS- and ELM-based timing synchronizers. With exhaustive experiments, the effectiveness of the proposed strategy is validated in channel scenarios of 5G (which are given in 3GPP TR 38.901 and have not been used for training), presenting its good generalization performance against the CIR uncertainty.
\vspace{-1mm}
\ifCLASSOPTIONcaptionsoff
  \newpage
\fi
   \vspace{-1.5mm}

\end{document}